\documentstyle[12pt]{article}
\begin{document}
\begin{center}
{\bf Punctuated equilibrium in an evolving bacterial population}
\vskip 1cm
Indranath Chaudhuri and Indrani Bose
\\Department of Physics, \\Bose Institute,
\\93/1, Acharya Prafulla Chandra Road,
\\Calcutta-700 009, India.
\end{center}

\begin {abstract}
Recently, Lenski et al have carried out an experiment on bacterial evolution.
Their findings support the theory of punctuated equilibrium in biological
evolution. We show that the M=2 Bak-Sneppen model can explain some of the 
experimental results in a qualitative manner.
\end{abstract}

There are two major schools of thought in evolutionary biology: gradualism
which implies continuous evolutionary changes and the theory of punctuated
equilibrium (PE) which states that evolutionary activity occurs in bursts.
Long periods of `stasis' are followed by short periods of rapid changes.
Recent exhaustive studies of fossil beds lend support to the second theory
\cite{R}. In a remarkable but controversial experiment, Lenski et al \cite{Elena,Lenski} studied
an evolving bacterial population for approximately 10,000 generations. They
inoculated a flask of low sugar broth with a dollop of bacteria. At the end 
of a day, a bit of the bacterial broth was siphoned into a fresh flask of
food to keep the cells growing and dividing. Every 15 days, a sample 
bacterial population was frozen for later analysis. After four years, data
for 10,000 generations were available. Lenski et al found evidence of PE
when they measured the average cell size every 100 generations. The relative
fitness, a measure of the increase in the growth rate of the descendant
population over that of the ancestral population, also increases in a step-like
manner. Further, the average cell size and the mean fitness appear to be 
correlated.

Lenski et al have interpreted the experimental results as suggesting that
natural selection of rare beneficial mutations may be responsible for the
punctuated growths seen in their experiments. During the periods of stasis,
the bacteria are in a sense waiting for the beneficial mutations to occur.
The rate of evolution is limited by the availability of such mutations.
The bacterial (E. coli) population was allowed to develop from a single
cell of a strain that can not exchange DNA. Spontaneous mutations are thus
the only source of genetic change in the population. The experiment also
involved competition for food because of the low availability of sugar.
This constitutes a selection pressure on the bacterial population. There
is a possibility that natural selection favours large cells as they have
more energy reserve. The extra biomass can lead to faster replication than
in the case of smaller cells. On the other hand, large cell size may simply
be correlated with other traits which are the targets of natural selection
and which are responsible for faster growth of the bacterial population.

Bak and Sneppen (BS) \cite{Bak} have earlier proposed a model of biological evolution
which exhibits PE in evolutionary activity in the so-called self-organised critical
(SOC) state. There is a modified version of the BS model known as the M-trait
BS model \cite{B,Bo} in which several biological species are considered, each of which
is characterised by M traits, instead of just one trait (fitness) as in the
original BS model. In this paper, we show that the M=2 BS model can explain
the major results of Lenski et al's experiment in a qualitative manner.

In the BS model, each site of a one-dimensional (1d) lattice represents a 
particular species. We make a slight modification to this model. We divide the 
bacterial population into $N$ categories. Each category contains bacteria of
similar characteristics. The $N$ categories correspond to the $N$ sites in the 
lattice. Two traits, namely, cell size and fitness (related to the replication
rate), are associated with the population at each site. The dynamics of evolution
is the same as in the M=2 BS model. One assigns a random number (between 0
and 1) to each of the traits at all the $N$ sites. At each time step, the site
with the minimum random value for a trait is identified. `Mutation' occurs to
bring about a change in the trait. The minimum random number is replaced by a 
new random number. The random numbers associated with any one of the traits of
the neighbouring sites are also replaced by new numbers. This is to take into
account the linkage of neighbouring populations in food chain. The last two
steps are repeated and averages are taken for both the traits locally (over 40
sites) and globally (over 2000 sites). Unlike in the original BS model, we
calculate quantities from the very beginning and not after the SOC state is
reached. The relative fitness (RF) is defined to be the ratio of current 
fitness and initial fitness at time t=0. In the actual experiment, fitness
is related to the growth rate of the bacterial population via replication.
In the following, we compare the results obtained by simulation with the 
experimental results of Lenski et al. Fig.1 shows RF versus time (generations)
for both experiment and simulation. In simulation, the RF is calculated for
the whole bacterial population by taking a global average over all sites.
Fig.2 shows the same curves but, now in the experiment, readings are 
taken every 100 generations, in contrast to 500 generations in Fig.1. The
step-like curve of the experiment is obtained in the simulation by taking a
local average over 40 sites in calculating the RF. The data points are 
obtained every 100 time steps. Figs.3 and 4 show results similar to those
in Figs.1 and 2 but now for the cell size (volume). Figs.1 and 3 show
that the RF and the average cell size increase quite rapidly in the first
2000 generations in the experimental environment. After several thousand
generations, changes are imperceptible. The data points in the two Figures
are taken every 500 generations. Figs.2 and 4 exhibit data taken every 100
generations. The step-like features correspond to periods of stasis followed
by short periods of rapid evolution. The time scale over which data is obtained
seems to be the deciding factor in the observation of PE. Punctuation shows
up only when the RF and the average cell size are measured every 100
generations. At the longer time scale of 500 generations, the growths appear
to be smooth and gradual. Objection has been raised \cite{Mlot} that if one
looks closely enough at anything, steps will eventually be seen. In the BS
model, gradual changes are seen when global averages are taken and steps
appear for local averages. Fig.5 shows the correlation between average cell
size and mean fitness for both experiment and simulation. The correlation
coefficients \cite{P} are r = 0.954 (experiment), r = 0.998 (simulation).

From Figs. 1-5, it appears that the experimental and simulation results
are in qualitative agreement. Many issues, however, have to be clarified
before a quantitative agreement can be obtained. As already mentioned,
Lenski et al have interpreted their experimental results by assuming that
natural selection favours beneficial mutations which confer competitive
advantage. In the BS model, mutations can be both beneficial and disadvantageous.
In the first case, the smallest fitness value is replaced by a larger value.
In the second case, the new fitness value is lower than the first one.
Simulation of the BS model, however, shows an overall increase of the RF
(or the cell size) in time. This is because the probability of beneficial bacterial
mutation is more in the BS model. In fact, in the SOC state, most fitness
values are above a critical threshold value. Consider that in a particular
time step, the minimal fitness value is 0.2. This number is to be replaced
by a random number drawn from a uniform distribution over the interval 0-1.
Thus the probability that the new fitness value is less than the original
one is 0.2 and the probability that the fitness increases is 0.8. Over a
large number of generations, the average fitness (or the cell size) has a
tendency to increase. Comparison of experimental and simulation results,
however, shows that the cell volume, obtained from simulation, has a much
smaller rate of growth than in the actual experiment. Also, the number of
steps in the cell size is larger than as seen in the experiment. The
experimental data for the RF (Fig.2) show four significant steps: 0 to
200, 300 to 500, 600 to 1200 and 1300 to 2000. The steps in the case of
the cell size (Fig.4) occur in the regions 0 to 200, 300 to 500, 600,
700 to 1100, 1200 and 1300 to 3000. Thus there are only two discrepancies
among the  step regions in the cases of the RF and the cell size. The
discrepancy is more in the case of simulation.

The relationship between simulation time steps and bacterial generations
is not clear. In the experiment, about $10^{6}$ mutations occurred every
day and there were approximately 6.6 generations per day. In the lattice
model, the bacterial population has a coarse-grained representation.
Bacteria of similar characteristics are grouped in a single category.
So one mutation in the lattice model is equivalent to a large number of
mutations in reality. In the experiment, serial transfer of a part of the
bacterial poulation to fresh sugar medium occured at the end of a day. This
feature is not explicit in the lattice model. Increase of the fitness
parameter indicates a higher replication rate and so the population
increase is taken into account in an indirect manner. One crucial feature
of the experiment is the scarcity of food (sugar). Again, this feature is
not inherent in the lattice model. The increase of the average cell size,
over time, in the lattice model can, however, be considered to be an
outcome of low food supply. The BS model has to be suitably modified to
take into account the effect of competition for food. Also, the implication
of the local versus the global average, in an actual experiment has to be
understood. The SOC state of the BS model is characterised by power-law
correlations \cite{Bak} in space and time. The growth of the RF and the
cell size, however, occur before the SOC state is reached. The SOC state
corresponds to the saturation regions in the growth curves. Bak and Sneppen
have shown that this state exhibits PE in the biological activity and
has the characteristic that all mutations involve changing of fitness
values below a self-organised critical value. Also, the distribution of
distances between successive mutations has a power-law decay as a function
of the distance between the mutations. The size
and lifetime distributions of avalanches of successive mutations have a
power-law form. Further experiments on bacterial evolution can test
the validity of some of these ideas.

\section*{Acknowledgement}
We thank Richard E. Lenski for permission to reproduce the original figure
in Refs. \cite{Elena} and \cite{Lenski}. One of the author I.C. is supported
by the Council of Scientific and Industrial Research, India under Sanction
No. 9/15(173)/96-EMR-I.

\newpage
\section*{Figure Captions}
\begin{description}
\item[Fig.1] Relative fitness versus time in experiment \cite{Lenski} and simulation.
A global average is taken over 2000 sites of the lattice to obtain the
data points in simulation.
\item[Fig.2] Relative fitness versus time in experiment \cite{Lenski} and simulation.
A local average is taken over 40 sites in simulation. the experimental data
points are taken every 100 generations.
\item[Fig.3] Average cell size versus time in experiment \cite{Lenski} and simulation
(global average).
\item[Fig.4] Average cell size versus time in experiment \cite{Elena} and simulation
(local average).
\item[Fig.5] Average fitness versus average cell size in experiment \cite{Elena} and
simulation.
\end{description}

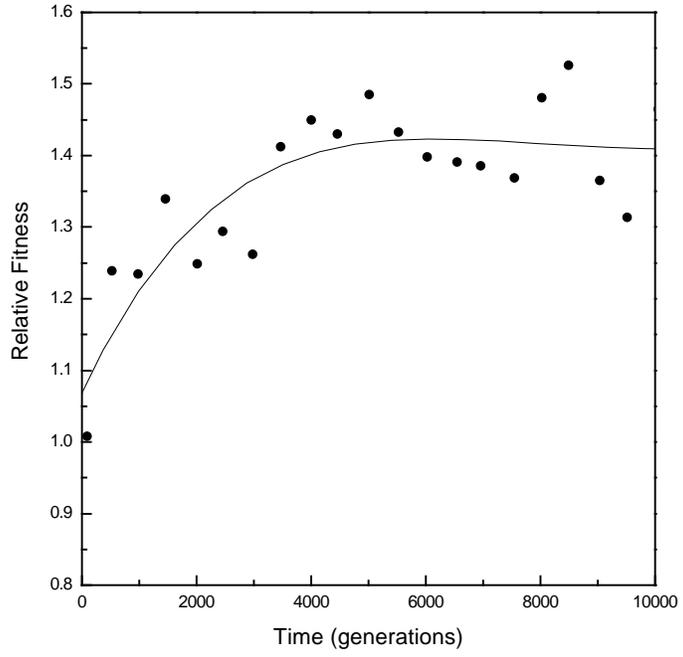
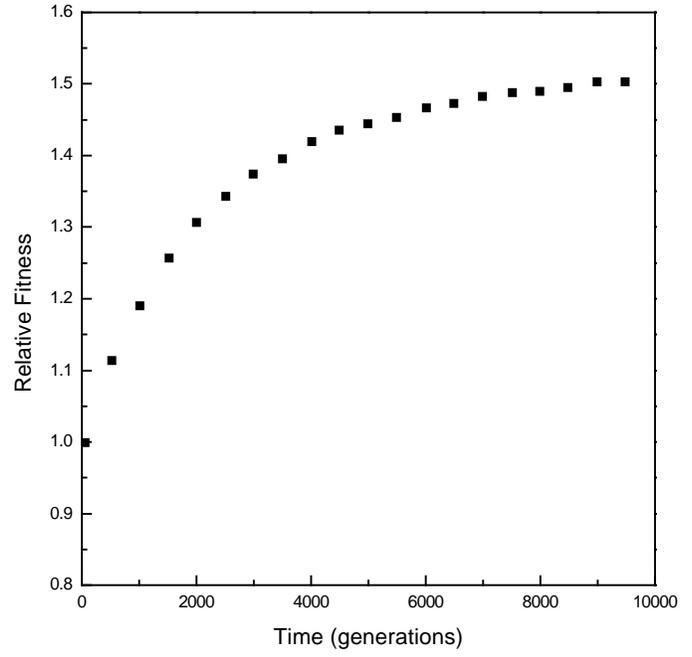
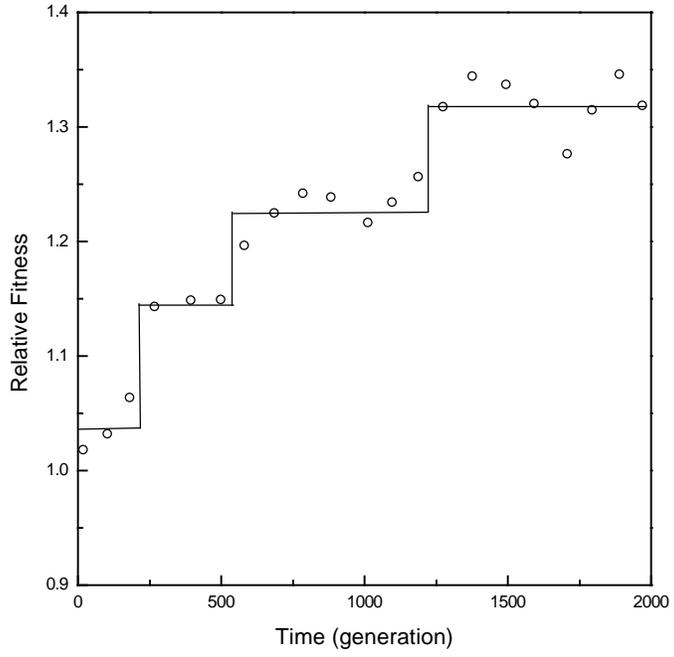
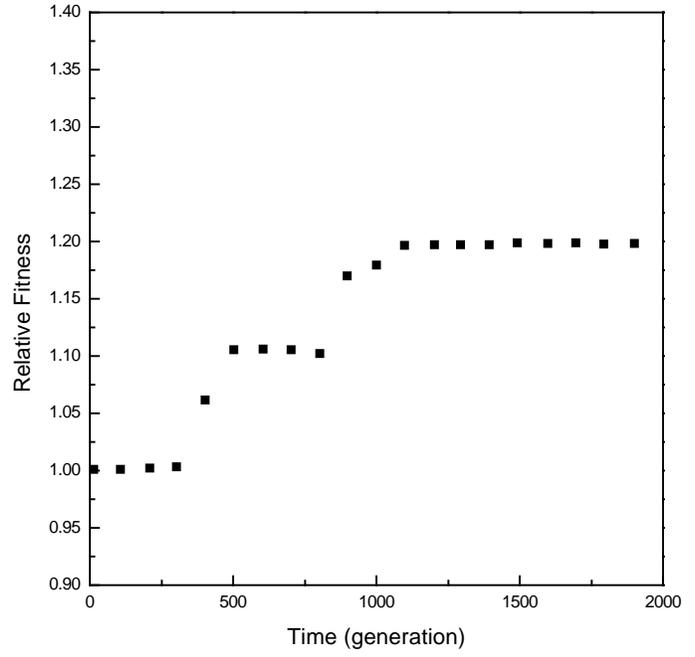
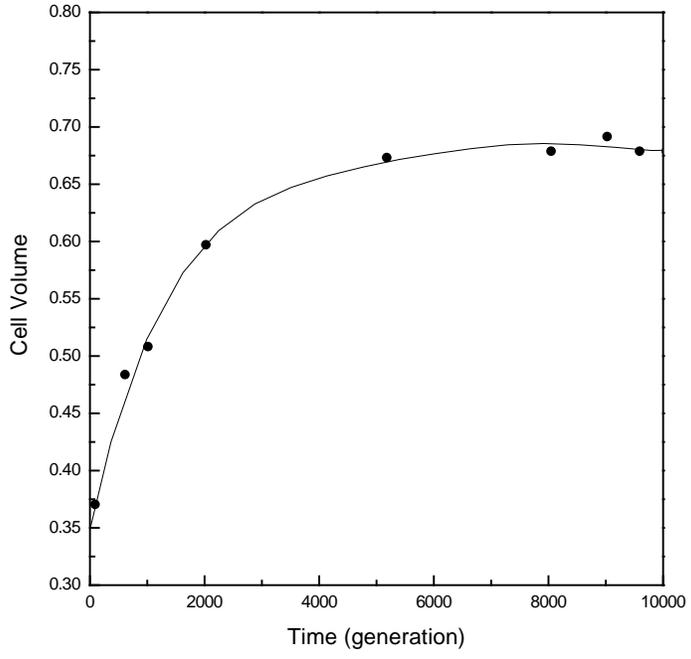
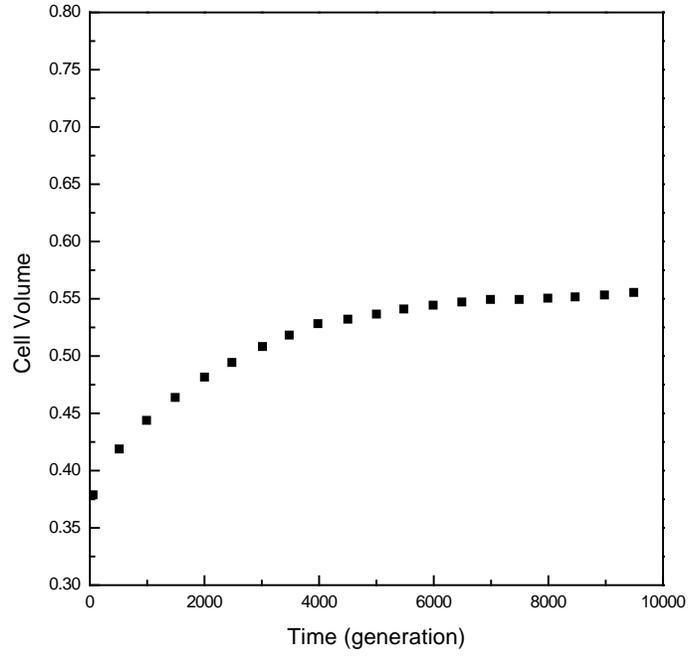
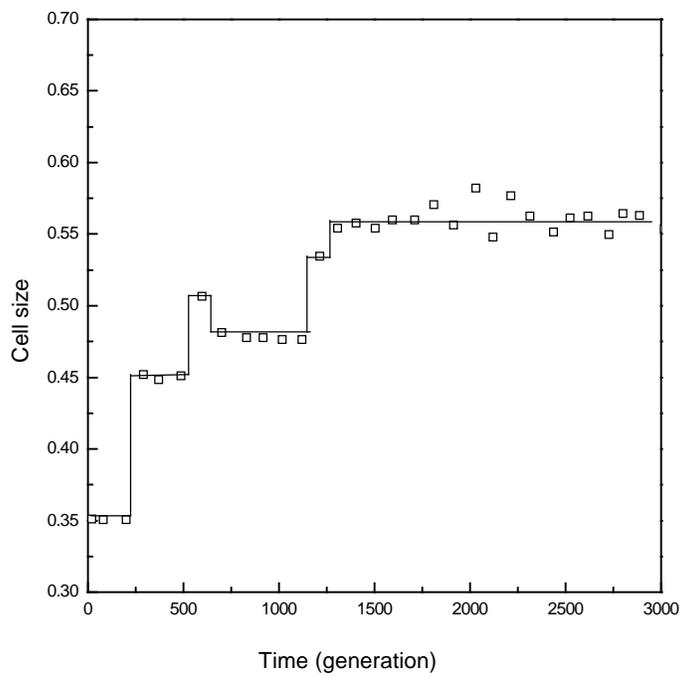
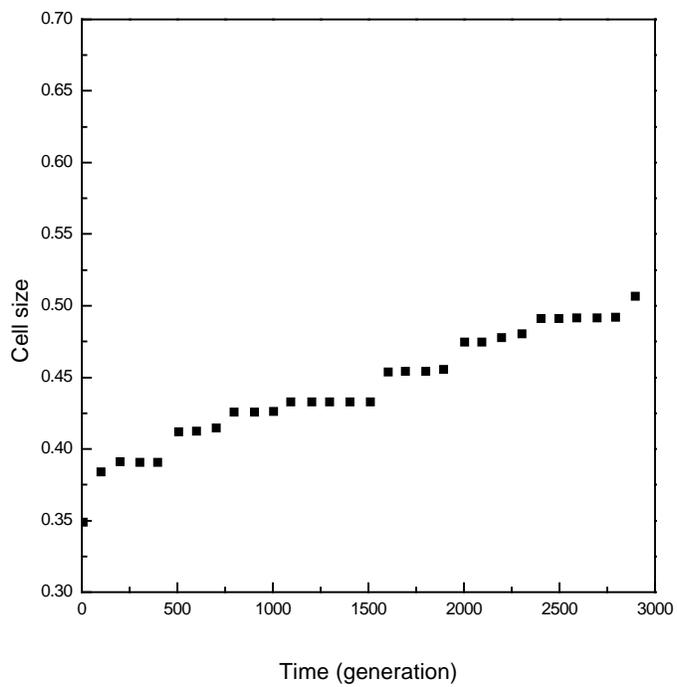
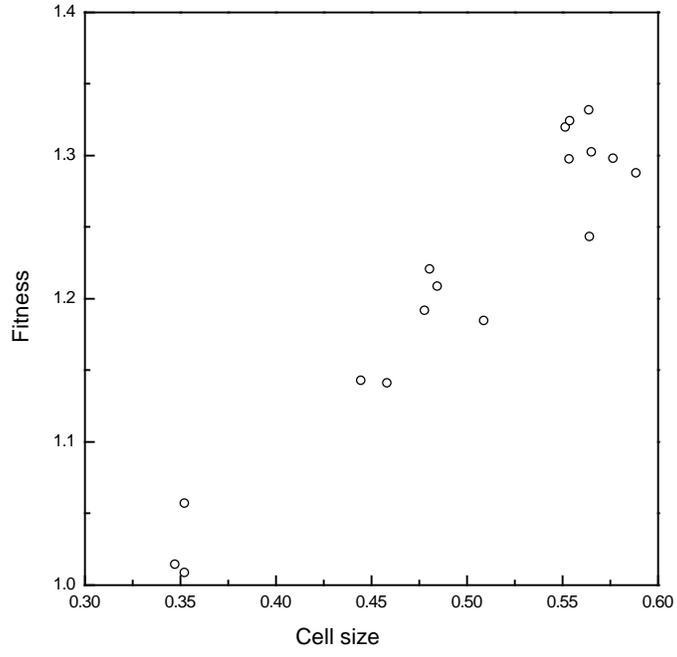
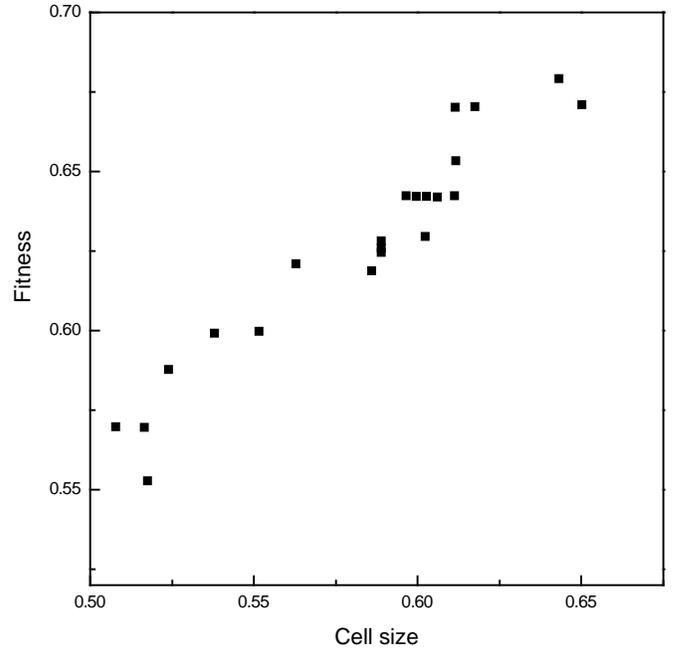
\end{document}